\documentclass[a4paper,twocolumn,english,aps, prd, superscriptaddress, preprintnumbers, showpacs, nofootinbib]{revtex4}
\usepackage[T1]{fontenc}
\usepackage[latin1]{inputenc}
\usepackage{amsmath}
\usepackage{amssymb}

\makeatletter
\usepackage{amssymb}

\usepackage{amsfonts}
\usepackage{latexsym}
\usepackage[T1]{fontenc}
\usepackage{ae,aecompl}

\usepackage{babel}

\usepackage{babel}
\makeatother
\begin{document}

\preprint{EPFL-ITS-14.2004; UCL-IPT-04-09}

\title{Cosmic microwave background constraints on the strong equivalence
principle}

\author{V. Boucher}

\author{J.-M. Gérard}

\affiliation{Institut de Physique Théorique, Université catholique de Louvain,
B-1348 Louvain-la-Neuve, Belgium}

\author{P. Vandergheynst}

\author{Y. Wiaux}

\email{yves.wiaux@epfl.ch}

\affiliation{Signal Processing Institute, Swiss Federal Institute of Technology,
CH-1015 Lausanne, Switzerland}

\date{July 2004}

\begin{abstract}
We study the effect of a violation of the strong equivalence principle
(SEP) on the cosmic microwave background (CMB). Such a violation would
modify the weight of baryons in the primordial gravitational potentials
and hence their impact in the establishment of the photon-baryon plasma
acoustic oscillations before recombination. This cosmological Nordtvedt
effect alters the odd peaks height of the CMB temperature anisotropy
power spectrum. A \emph{gravitational baryonic mass density} of the
universe may already be inferred at the first peak scale from the
analysis of WMAP data. Experimental constraints on a primordial SEP
violation are derived from a comparison with the universe's \emph{inertial
baryonic mass density} measured either in a full analysis of the CMB,
or in the framework of the standard big bang nucleosynthesis (BBN). 
\end{abstract}


\maketitle

\section{Introduction}

The recent results of the cosmic microwave background (CMB) experiments,
together with other cosmological tests, provide us today with a coherent
picture of the structure and evolution of the universe. The corresponding
canonical paradigm postulates a spatially flat universe which has
undergone a period of inflation in its early ages. The present large
scale structure of the universe essentially originates from primordial
quantum energy density fluctuations around a homogeneous and isotropic
background. These perturbations also left their imprint on the cosmic
background radiation which decoupled from the rest of the universe
some $380.000$ years after the big bang. According to this cosmological
model, the universe is filled in with about seventy percent of dark
energy, twenty-five percent of cold dark matter, five percent of ordinary
(baryonic) matter and a relic background of radiations. The recent
one-year WMAP results led to an already precise determination of the
corresponding cosmological parameters \cite{page,spergel}. However,
in this context full credit may not be given to this analysis before
the theoretical hypotheses on which the canonical paradigm is based
are tested, notably through a thorough analysis of the CMB. Many questions
may in fact be raised. The inflationary scenario and the structure
of the initial conditions for energy density perturbations have been
extensively analyzed (see for example \cite{bouchet,coles} for general
considerations). Lately, the well established cosmological principle,
postulating global homogeneity and isotropy of the universe, has also
been challenged \cite{coles,hansen,hajian}. But, perhaps most fundamentally,
one should question the theory of gravitation on which cosmology is
developed, namely general relativity.

Any explicit theory of gravitation beyond general relativity introduces
different effects which modify the characteristics of the CMB temperature
anisotropy power spectrum. In particular the location and height of
acoustic peaks should be altered proportionally to the strength of
auxiliary gravitational couplings. This analysis has already been
performed for pure Brans-Dicke theories \cite{chen} as well as generalized
versions \cite{catena,nagata2,nagata1}, leading to constraints on
a possible scalar coupling. However these bounds do not take into
account possible effects induced by the violation of the strong equivalence
principle (SEP). This principle, essential feature of the theory of
general relativity, notably postulates the constancy of the newtonian
gravitational coupling $G$ in space and time. It distinguishes Einstein's
theory from other metric theories of gravitation. Any SEP test therefore
challenges general relativity in its most fundamental structure. The
purpose of this article is to study the influence of a possible SEP
violation on the CMB temperature power spectrum. In order to single
out the generic effect of such a SEP violation on the CMB, our approach
naively conserves Einstein's equations for the gravitational field.
Hence, under the hypothesis of the cosmological principle, the Friedmann-Lemaître
background and perturbed cosmological evolution remains unchanged.
In this way, we avoid the introduction of multiple effects which could
unnecessarily complicate the conceptual analysis. The SEP violation
is simply introduced through the break down of energy-momentum conservation
for compact bodies. The breaking term in the corresponding covariant
equations depends on the gradient of the gravitational coupling $G$
with respect to spacetime coordinates.

In section \ref{sec:SEP} we introduce the SEP and its violation.
Section \ref{sec:SEP-and-CMB} is devoted to the analysis of the effect
of a SEP violation on the CMB temperature power spectrum. The SEP
violation affects the weight of baryons in the primordial gravitational
potentials and hence their impact in the establishment of acoustic
oscillations of the plasma before the last scattering of photons.
The odd peaks height enhancement of the CMB temperature power spectrum
depends indeed on a gravitational baryonic mass density of the universe,
not on its inertial baryonic mass density. We also discuss the characteristic
amplitude of this effect and the uniqueness of its signature relatively
to the variation of the canonical cosmological parameters. In section
\ref{sec:EC} we derive experimental constraints on the SEP violation.
A gravitational baryonic mass density of the universe is inferred
at the first peak scale from WMAP data. The constraints on the SEP
violation are obtained from the measurement of the inertial baryonic
mass density, either in a full analysis of the CMB temperature power
spectrum, or through the independent determination of light element
abundances in the framework of standard BBN. We discuss the proposed
constraints, and finally conclude.

This article pursues the work done in \cite{boucher}, on the ground
of both theory and data analysis.

\section{SEP violation \label{sec:SEP}}

The equivalence principle is an important fundament of any theory
of gravitation. It is however implemented at different levels in different
theories. The Einstein equivalence principle postulates the universality
of free fall of test-bodies at one given point of a gravitational
field (called weak equivalence principle), as well as the independence
of the result of any non-gravitational experiment in a freely falling
frame relative to the velocity of free fall and relative to where
and when in the universe it is performed. A metric theory of gravitation
postulates the geodesic motion of test-bodies, as well as the agreement
of the results of any non-gravitational experiment performed in free
fall with the laws of special relativity. By definition, all metric
theories of gravitation therefore respect the Einstein equivalence
principle. Technically, a theory of gravitation respects this principle
if the Lagrange density for matter only depends on the matter fields
and the spacetime metric, but not on possible auxiliary gravitational
fields which directly couple to the metric. This structure implies
indeed the general covariant conservation equations $T_{\quad\shortmid\nu}^{\mu\nu}=0$
for the energy-momentum tensor $T^{\mu\nu}$, from which readily follows
geodesic motion. On the other hand, non-gravitational interactions
are coupled to the metric field through the connexion, and therefore
reduce to their special-relativistic structure in free fall. Both
postulates of the Einstein equivalence principle are therefore ensured.

The strong equivalence principle extends the universality of free
fall to compact bodies. By compact body, one means a body with a non-negligible
amount of internal gravitational binding energy. It also extends to
gravitational experiments the independence of the result of any experiment
in free fall relative to the velocity of free fall and relative to
where and when in the universe it is performed. The mere existence
of an auxiliary field of gravitation coupled to the metric field violates
this principle. The reason for this holds in the fact that it is not
possible to cancel the effect of auxiliary (typically scalar or vector)
fields by a local coordinate transformation, like it is for the tensor
metric field. Auxiliary couplings will inevitably modify the result
of gravitational experiments (and notably the structure of compact
bodies) performed in a freely falling frame, therefore violating the
SEP. The gravitational coupling $G$ itself will depend on the spacetime
point through a dependence in the auxiliary fields, which implies
by definition to a SEP violation. Aside from the Nordström scalar
theory, only general relativity incorporates the equivalence principle
at the level of the SEP \cite{gerard,damour1}. Testing the SEP violation
is therefore a way of discriminating general relativity from other
metric theories of gravitation such as extended Brans-Dicke or vector-tensor
theories of gravitation.

If the newtonian gravitational coupling is a function of the position
$x$ in spacetime, $G\rightarrow G(x)$, the mass $m$ of a compact
body also depends on the position through its internal gravitational
binding energy. An effective action for the geodesic motion of compact
bodies may therefore be defined as: $S_{mat}=-c\int m(x)ds$. Energy-momentum
conservation is therefore broken through the introduction of a source
term in the general covariant conservation equations. We adopt the
corresponding expression as our mathematical implementation of a possible
SEP violation:\begin{eqnarray}
T_{\quad\shortmid\nu}^{\mu\nu} & = & G^{,\mu}\frac{\partial T}{\partial G}\quad,\label{eq:SEP1}\end{eqnarray}
 where $T$ is the trace of the energy-momentum tensor. The dependence
of the newtonian gravitational coupling on the spatial position $\vec{x}$
is parametrized through the relation $G(\vec{x})=G_{0}(1+\eta_{g}V(\vec{x})/c²)$,
where $V(\vec{x})$ stands for the gravitational potential at the
point considered, $G_{0}$ is the background value of the newtonian
gravitational coupling in the absence of this potential, and $\eta_{g}$
is the parameter which defines the amplitude of the SEP violation.
One may also define the compactness $s$ of a body as the sensitivity
of its mass relative to $G$. It is equivalently given by the ratio
of its internal gravitational binding energy $E_{g}$ to its total
mass energy: $s=-d\ln m/d\ln G=|E_{g}|/mc^{2}$. The acceleration
$\vec{a}$ of a body in a gravitational field now explicitly depends
on its proper sensitivity $s$. This establishes the SEP violation
through the so-called Nordtvedt effect \cite{nordtvedt1,nordtvedt2}.
From the definition (\ref{eq:SEP1}) indeed, we get in the non-relativistic
(called quasi-newtonian) approximation: $\vec{a}=\vec{g}(1-\eta_{g}s)$,
where $\vec{g}=-\vec{\nabla}V(\vec{x})$. In other words, the SEP
violation induces a reduction (for $\eta_{g}>0$) or increase (for
$\eta_{g}<0$) of the gravitational mass $m_{g}$ of a body relative
to its inertial mass $m$, proportionally to its own compactness:
\begin{eqnarray}
m_{g} & = & m\left(1-\eta_{g}s\right)\quad.\label{eq:SEP1-1}\end{eqnarray}

From the experimental point of view, tests of the weak equivalence
principle date back to Newton and its pendulum experiments. First
tests of the SEP have been introduced several decades ago with the
Lunar Laser Ranging experiment (see \cite{will} for an extended review
and references). This test still gives the best constraint on the
parameter $\eta_{g}$ today%
\footnote{Quantities measured at the present epoch are indexed by the superscript
$^{0}$.%
}:\begin{eqnarray}
\eta_{g}^{0} & \leq & 1\times10^{-3}\quad,\label{eq:SEP-1-2}\end{eqnarray}
 with involved compactnesses of order $10^{-10}$ \cite{will,williams}.
A better constraint may be inferred in the framework of peculiar scalar-tensor
theories of gravitation though. Indeed, the SEP violation actually
introduces a new charge of gravitation beyond the mass, in terms of
the compactness $s$. This new charge not only modifies the motion
of compact bodies (Nordtvedt effect) but affects the dynamical structure
of the corresponding theory, inducing potentially dominant dipole
gravitational radiations associated with auxiliary fields of gravitation
(scalar or vector). The analysis of the orbital period decrease rate
of asymmetric binary pulsars is an extremely good probe of dipole
radiations. The recent and unique measurement of the orbital decrease
in such a binary, the neutron star - white dwarf $\textnormal{PSR\, J}1141-6545$
\cite{bailes,kaspi} gives a tight constraint on a scalar gravitational
coupling. In pure Brans-Dicke theories (BD), the corresponding bound
on the SEP violation inferred from \cite{gerard,bailes} reads: $\eta_{g}^{0(BD)}\leq2.7\times10^{-4}$
(see also \cite{esposito}).

\section{SEP and CMB\label{sec:SEP-and-CMB}}

\subsection*{Qualitative analysis}

In the primordial universe the photon gas is rather tightly coupled
to electrons through Compton scattering. The electrons are themselves
linked to protons through the Coulomb interaction. We may then consider
a photon-baryon plasma in evolution in gravitational potentials. These
gravitational potentials are essentially produced by the dominant
cold dark matter component of the universe. About $380.000$ years
after the big bang, the temperature of the expanding universe had
decreased too much to longer maintain hydrogen dissociation. The cosmic
microwave background radiation observed today corresponds to a snapshot
of this photon gas which decoupled from the rest of the universe at
the time of last scattering. The anisotropy distribution on the sky
today is determined by the multiple physical phenomena which governed
the evolution of the plasma before recombination, and therefore contains
all the information on the structure and evolution of the universe
(defined in terms of cosmological parameters). The plasma underwent
oscillations, responsible for relative temperature fluctuations in
the associated black-body spectrum. In the corresponding angular power
spectrum, this oscillation process translates into a series of acoustic
peaks at scales smaller than the horizon size at last scattering.
Odd peaks correspond to scales which had reached maximum compression
(rarefaction) at the time of last scattering in potential wells (hills).
Even peaks correspond to maximum rarefaction (compression) in potential
wells (hills). The general shape of this spectrum therefore exhibits
a Sachs-Wolfe plateau at scales beyond the horizon size at last scattering,
followed by the acoustic peak series under the horizon size. Notice
that, up to now, the standard CMB analysis has been based on the study
of the precise characteristics of the temperature anisotropy angular
power spectrum. The cosmological parameters are determined through
a best fit of the theoretical cosmological models with experimental
data (see notably \cite{page,spergel} for the WMAP analysis).

The oscillations of the plasma are electromagnetic acoustic oscillations
of the photon gas. However, the action of gravity is introduced through
a purely newtonian coupling of the baryonic content of the plasma
to the dark matter potentials. The effect of this coupling is to shift
the zero point (equilibrium) of the oscillations toward more compressed
states in potential wells, and rarefied states in potential hills.
Consequently, the height of odd peaks relative to even peaks is enhanced
proportionally to the total baryon weight in the dark matter potentials
\cite{hu01,hu96,hu95a,hu95b,hu95c}.

If the SEP is violated through a spatial dependence of the newtonian
gravitational coupling (Nordtvedt effect), gravitational masses differ
from inertial masses. The temperature power spectrum peaks height
therefore bears the imprint of a possible SEP violation as it essentially
originates from a gravitational interaction and therefore depends
on a gravitational baryonic mass density:\begin{eqnarray}
\left(\rho_{b}\right)_{g} & = & \rho_{b}\left(1-\eta_{g}s_{b}\right)\quad,\label{eq:SEP-CMB-1-0}\end{eqnarray}
 rather than on the inertial baryonic mass density $\rho_{b}$. The
compactness $s_{b}$ must be associated with a baryon-region seen
as a compact body at the relevant cosmological scale.

\subsection*{Plasma evolution equations and SEP violation }

The purpose of this subsection is to derive more technically the main
result of the last subsection. The evolution equations for the photon-baryon
plasma in the tight coupling limit are derived from the generalized
covariant energy-momentum tensor equations (\ref{eq:SEP1}).

The tight coupling limit amounts to consider an infinite Compton interaction
rate which implies the equality of the mean photon and baryon velocities:
$\vec{v}_{\gamma}=\vec{v}_{b}$. In this standard approximation, the
photon-baryon gas may be entirely described as a fluid with the energy-momentum
tensor $T^{\mu\nu}=(\rho+P/c^{2})u^{\mu}u^{\nu}-Pg^{\mu\nu}$. The
equation of state relating pressure and density reads: $P=\lambda\rho c^{2}$,
with $\lambda=0$ for matter and $\lambda=1/3$ for radiation. Restricting
ourselves to a flat universe, in the newtonian gauge, with conformal
time $\eta$ and comobile coordinates $\vec{x}$, we may write the
perturbed spacetime metric as $g_{00}(\vec{x},\eta)=a^{2}(\eta)(1+2\Psi(\vec{x},\eta)/c^{2})$,
$g_{0i}(\vec{x},\eta)=0$, and $g_{ij}(\vec{x},\eta)=-\delta_{ij}a^{2}(\eta)(1+2\Phi(\vec{x},\eta)/c^{2})$.
The factor $a(\eta)$ stands for the scale factor of the expanding
universe normalized to its present size ($a^{0}=1$). The scalar perturbations
$\Psi(\vec{x},\eta)$ and $\Phi(\vec{x},\eta)$ may been seen as newtonian
potentials.

From the equations (\ref{eq:SEP1}) we readily obtain the continuity
and Euler equations for the fluid under consideration. In the Fourier
space, to first order in the relative density perturbations $\delta(\vec{k},\eta)$,
comobile velocity $v(\vec{k},\eta)$, and gravitational potentials
$\Psi(\vec{k},\eta)$ and $\Phi(\vec{k},\eta)$, these equations read
respectively:\begin{eqnarray}
\dot{\delta} & = & -\left(1+\lambda\right)\left(i\vec{k}\cdot\vec{v}+3\frac{\dot{\Phi}}{c^{2}}\right)-\left(1-3\frac{c_{s}^{2}}{c^{2}}\right)s\,\frac{\dot{G}}{G}\label{eq:SEP-CMB1}\\
\dot{\vec{v}} & = & -\frac{\dot{a}}{a}\left(1-3\frac{c_{s}^{2}}{c^{2}}\right)\vec{v}-i\vec{k}\left[c_{s}^{2}\frac{\delta}{1+\lambda}\right.\nonumber \\
 &  & \left.+\Psi\left(1-\left(1-3\frac{c_{s}^{2}}{c^{2}}\right)\frac{\eta_{g}s}{1+\lambda}\right)\right]\quad.\label{eq:SEP-CMB2}\end{eqnarray}
 Dotted variables here stand for their derivative with respect to
the conformal time. The sound speed in the fluid $c_{s}$ and the
compactness $s$ characterizing a given fluid volume are background
space-independent quantities. The $s$-terms represent the explicit
modification due to SEP violation of the canonical \cite{hu95a,hu95b,hu95c}
evolution equations for a single component fluid.

In order to find the evolution equations for the photons, we just
apply this set of equations to a photon fluid with $\vec{v}=v_{\gamma}\hat{k}$,
taking into account the presence of baryons in the sound speed and
the compactness. The sound speed reads $c_{s}^{2}=dP_{\gamma}/d(\rho_{\gamma}+\rho_{b})=c^{2}/3(1+R)$,
where $\rho_{b}$ and $\rho_{\gamma}$ are respectively the background
inertial baryonic mass density and photon density of the universe,
and $R=3\rho_{b}/4\rho_{\gamma}$ is the canonical normalization of
the baryonic mass density by the photon density. The photon gravitational
binding energy is negligible and the fluid compactness reduces to
the baryonic component $s_{b}$, which is studied in the next subsection.
The fluid density and velocity may be expressed in terms of the monopole
and dipole moments $\Theta_{0}(\vec{k},\eta)$ and $\Theta_{1}(\vec{k},\eta)$
of the photon relative temperature distribution: $\delta_{\gamma}(\vec{k},\eta)=4\Theta_{0}(\vec{k},\eta)$
and $v_{\gamma}(\vec{k},\eta)=-3i\Theta_{1}(\vec{k},\eta)$. In this
context, the plasma evolution equations for $X(\vec{k},\eta)=\Theta_{0}(\vec{k},\eta)+\Phi(\vec{k},\eta)/c^{2}$
and $\Theta_{1}(\vec{k},\eta)$ read:\begin{eqnarray}
\ddot{X}+\frac{\dot{R}}{1+R}\dot{X}+k^{2}c_{s}^{2}X & = & k^{2}c_{s}^{2}\left[\frac{\Phi}{c^{2}}\right.\nonumber \\
 &  & \left.-\frac{\Psi}{c^{2}}\left(1+R\left(1-\eta_{g}s_{b}\right)\right)\right]\label{eq:SEP-CMB3}\\
k\Theta_{1} & = & -\dot{\Theta}_{0}-\frac{\dot{\Phi}}{c^{2}}\quad.\label{eq:SEP-CMB3-1}\end{eqnarray}
 We do not consider here the term with temporal dependence of the
newtonian coupling, though it would be worth analyzing its effect.
Only the spatial dependence of $G$ is considered by analogy with
the Nordtvedt effect. The first equation sets the dynamics for damped
oscillations for $\Theta_{0}$ with a forcing term (right-hand side).
We clearly identify that the effect of baryons in this forcing term
depends indeed on the gravitational baryonic mass density\begin{eqnarray}
R_{g}\left(s_{b},\eta_{g}\right) & = & R\left(1-\eta_{g}s_{b}\right)\quad,\label{eq:SEP-CMB4}\end{eqnarray}
 function of the compactness $s_{b}$, rather than on the inertial
baryonic mass density.

Notice that in the limit of constant newtonian potentials%
\footnote{Quantities measured at recombination are indexed by the superscript
$^{*}$.%
} $\Psi=\Psi^{*}$, $\Phi=\Phi^{*}$, with $R=R^{*}$, equation (\ref{eq:SEP-CMB3})
reduces, for the effective temperature perturbation $Y=\Theta_{0}+\Psi^{*}/c^{2}$,
to $\ddot{Y}+k^{2}c_{s}^{2}Y=-k^{2}c_{s}^{2}R_{g}^{*}(s_{b}^{*},\eta_{g}^{*})\Psi^{*}/c^{2}$.
The forcing term clearly reduces to the (quasi-)newtonian interaction
between the baryons and the surrounding constant potentials. In the
further approximations $s_{b}=s_{b}^{*}$ and $\eta_{g}=\eta_{g}^{*}$
discussed in the following, the interaction term is constant. We therefore
recover the exact limit in which a constant zero-point shift of the
acoustic oscillations originates the odd peaks height enhancement
of the temperature power spectrum. But the acceleration of baryons
is now a function of the compactness of the baryon-region considered.
Equation (\ref{eq:SEP-CMB4}) is therefore the mathematical expression
of the cosmological Nordtvedt effect discussed in the former qualitative
analysis.

\subsection*{Compactness of baryon-regions}

Under the hypothesis of the cosmological principle, we live in a globally
homogeneous and isotropic universe. As suggested in our qualitative
analysis, let us consider a homogeneous spherical baryon-region of
radius $L$ and total mass $M_{b}$. Its compactness calculated, in
the spirit of the quasi-newtonian approach introduced in section \ref{sec:SEP},
as the ratio of the internal gravitational binding energy over the
total mass energy reads: $s_{b}=3GM_{b}/5Lc^{2}=4\pi G\rho_{b}L^{2}/5c^{2}$.
The mean baryon density scales as $\rho_{b}(a)=\rho_{b}^{0}a^{-3}$.
At each instant in the course of the universe expansion, the maximum
size of the radius $L$ is set by the event horizon: $L_{1}\left(\eta\right)=ca\eta$.
This hypothesis is natural as the event horizon defines at each moment
the maximal distance through which particles may have interacted gravitationally
since the primordial ages of the universe (after inflation), and therefore
the maximal size of a body. In matter and radiation universes, the
Friedmann-Lemaître equations (in the considered limit where Einstein
equations are preserved) determine the evolution of the scale factor
with time as $\eta/\eta^{0}=a^{1/2}$ and $\eta/\eta^{0}=a$, respectively.

The compactness of a baryon-region therefore grows linearly with the
scale factor in a radiation era, while it is constant in a matter
era. Recombination takes place after the matter-radiation equilibrium,
inside the matter era. For the sake of the analogy with the Nordtvedt
effect on compact bodies in a gravitational field, we consider in
the following a constant compactness over the course of the universe
evolution until recombination. It is evaluated at its value in the
matter era, say at last scattering ($s_{b}=s_{b}^{*}$). The low baryon
density turns out to be largely compensated by the considered cosmological
scales to give a non-negligible contribution to the compactness. In
terms of physical quantities (the Hubble constant, the age of the
universe and the relative baryon density), we get a compactness \begin{eqnarray}
s_{b}^{1*} & = & \frac{27}{10}\left(H^{0}t^{0}\right)^{2}\Omega_{b}\simeq0.1\quad,\label{eq:SEP-CMB-5}\end{eqnarray}
 for the maximal radius $L_{1}$. This compactness is the sensitivity
to be considered at the scale of the wavelength $\lambda_{1}$ associated
with the first acoustic peak. The sensitivity of the baryonic body
relevant for the subsequent acoustic peaks ($\lambda_{n}$) scales
like $n^{-2}$ since the compactness $s_{b}$ of the baryon-region
considered is proportional to the square of its radius $L$: \begin{eqnarray}
s_{b}^{n*} & \simeq & 0.1n^{-2}\quad.\label{eq:SEP-CMB-6}\end{eqnarray}
Let us now briefly comment on the implications of these results.

\subsection*{Amplitude of the SEP violation effect}

The value $s_{b}^{1*}\simeq0.1$ in (\ref{eq:SEP-CMB-5}) implies
that a SEP violation parameter of order unity at the time of recombination,
$\eta_{g}^{*}\simeq1$, would affect the first peak height by $10$\%
(see equation (\ref{eq:SEP-CMB4})). In present CMB analyses, the
cosmological parameter $\Omega_{b}h^{2}$ identifying the baryon content
of the universe is essentially extracted from the measurement of the
relative height between the first and second peaks of the temperature
angular power spectrum. In this regard, it measures the gravitational,
rather than inertial, baryonic mass density of the universe. The recent
one-year WMAP analysis gives this parameter with a precision of $4$\%.
Consequently, the present CMB data will already allow us to derive
interesting constraints on a possible SEP violation.

\subsection*{Uniqueness of the SEP violation signature}

The peculiar $n^{-2}$ scaling of the baryon-regions compactness $s_{b}^{n*}$
in (\ref{eq:SEP-CMB-6}) ensures the orthogonality of the SEP violation
signature relative to the effect other cosmological parameters on
the CMB temperature angular power spectrum. The signature of the SEP
violation may indeed be disentangled from the effect of other parameters
through the corresponding $n^{-2}$ scaling of the odd peaks height.
The measurement of the SEP violation parameter $\eta_{g}^{*}$ at
recombination is therefore in principle possible, simultaneously to
the determination of the canonical \cite{hu01,kosowsky} cosmological
parameters. The Planck satellite is designed to achieve a better sensitivity
in the temperature anisotropies measurement, as well as a better resolution
on the sky, than the present WMAP mission. This mission will notably
give access to the whole series of acoustic peaks in the temperature
anisotropies angular power spectrum \cite{bouchet}, therefore allowing
an unambiguous analysis of a possible SEP violation.

\section{Experimental constraints\label{sec:EC}}

In this section, we establish experimental constraints on the SEP
violation parameter $\eta_{g}^{*}$ at recombination, and discuss
their significance in comparison with existing bounds at our epoch
and theoretical predictions at the exit of the radiation era.

A precise analysis of a possible SEP violation must be performed through
a best fit of our modified theory (\ref{eq:SEP1}) and experimental
data, taking into account the substitution (\ref{eq:SEP-CMB4}) in
the plasma evolution equations before recombination. Here, we determine
bounds on a possible SEP violation by the analysis of the one-year
WMAP experimental error bars on the observables of interest. This
simple approach finds its justification in the fact that our modified
theory assumes the cosmological Nordtvedt effect to be the only perturbation
to the cosmic background anisotropy spectrum relatively to the canonical
paradigm based on general relativity. In a first approach, one can
determine the gravitational and inertial baryonic mass densities of
the universe at recombination from their specific (orthogonal) signatures
on the CMB power spectrum characteristics. A second generic approach
consists in determining the gravitational baryonic mass density through
the analysis of the CMB, using as a prior the measurement of the inertial
baryonic mass density by independent observations. In that respect,
we will consider here the determination of the inertial baryonic mass
density through the measurement of light element abundances in the
framework of standard BBN.

\subsection*{CMB-CMB constraint}

The baryon content of the universe affects the CMB temperature power
spectrum in different ways. The major effect is a dependence of the
odd peaks height due to the weight of baryons in the surrounding gravitational
potentials. We already know that this effect is actually a function
of a gravitational baryonic mass density $R_{g}^{*}$. It bears the
imprint of a possible SEP violation in terms of the already discussed
$n^{-2}$ scaling. This unique signature adds to the canonical odd
peaks height enhancement related to the inertial baryonic mass density
$R^{*}$ (equation (\ref{eq:SEP-CMB4})). But any increase of the
baryon density also naturally induces a decrease of the sound speed
for the propagation of the acoustic oscillations in the primordial
plasma, therefore affecting the peaks location, rather than their
height. Increasing the baryon density also decreases the diffusion
length, defined as the scale below which inhomogeneities are damped
because of the finite Compton interaction rate. These last two effects
are related to electromagnetic (rather than gravitational) phenomena
and are consequently independent of the SEP violation. They only depend
on the inertial baryonic mass density $R^{*}$.

As already mentioned, the forthcoming Planck mission will probe all
these signatures. At present however, the temperature power spectrum
characteristics are known with precision only up to the second peak
through the one-year WMAP data. It is therefore rather difficult to
disentangle a SEP violation from variations of other cosmological
parameters, notably from $R^{*}$. However, assuming that all parameters,
other than $R_{g}^{*}$ and $R^{*}$, are fixed to their accepted
value, we may infer a constraint on $\eta_{g}^{*}$. On the one hand,
we consider the one-year WMAP value of the cosmological parameter
$\Omega_{b}h^{2}$ as a measure of the relative height between the
first and second peaks \cite{page}, hence originating from the gravitational
baryonic mass density $R_{g}^{*}(s_{b}^{1*},\eta_{g}^{*})$, at last
scattering, and at a scale corresponding to the maximum oscillation
wavelength. On the other hand however, the specific analysis of the
first peak position gives the inertial baryonic mass density $R^{*}$,
through the dependence of the peaks location in the sound speed in
the primordial plasma. A simple analysis of the one-year WMAPext (i.e.
WMAP extended to the CBI and ACBAR experiments \cite{page,spergel})
error bars on these two observables gives the bound: $|\eta_{g}^{*}s_{b}^{1*}|\leq0.06$.
From the estimated value (\ref{eq:SEP-CMB-5}) for $s_{b}^{1*}$ ,
we readily obtain the following constraint on the SEP violation in
terms of $\eta_{g}^{*}$:\begin{eqnarray}
|\eta_{g}^{*(CMB)}| & \leq & 0.6\quad.\label{eq:EC-1}\end{eqnarray}

\subsection*{CMB-BBN constraint}

The standard BBN model may also infer the inertial baryonic content
of the universe from the determination of light element ($D$, $^{3}He$,
$^{4}He$, $^{7}Li$) abundances. These abundances are studied in
low-metallicity systems in such a way that they still significantly
reflect primordial quantities. In this context, the baryon content
of the universe is usually quoted in terms of $\Omega_{b}h^{2}$,
rather than $R^{*}$. Notice that BBN is also affected in the framework
of a specific alternative theory of gravitation \cite{catena,carroll,damour2}.
However it is independent of the SEP violation considered here.

The primordial $^{4}He$ abundance is determined to better accuracy
than in the case of other light elements \cite{izotov,luridiana,fields}.
However, it is rather insensitive to the baryon content. The measurement
of $^{4}He$ abundance therefore has to be extremely precise if one
wants to obtain a small uncertainty on $\Omega_{b}h^{2}$ or $R^{*}$.
The most recent estimate, obtained in the analysis of dwarf irregular
and compact blue galaxies gives, for the $^{4}He$ mass fraction:
$Y_{p}=0.2421\pm0.0021$ \cite{izotov}. In the framework of the standard
BBN theory, the corresponding baryon content is $\Omega_{b}h^{2}=12_{-2}^{+3}\times10^{-3}$
or $R^{*}=0.334_{-0.056}^{+0.084}$. The baryon density inferred from
the primordial lithium-to-hydrogen abundance ratio $^{7}Li/H$ lies
around the same values \cite{ford,Bonifacio,ryan}. The one-year WMAPext
value, still understood as a measurement of the relative height of
the first two peaks of the CMB temperature power spectrum, gives a
significantly higher value for $R_{g}^{*}$: $\Omega_{b}h^{2}=(22\pm1)\times10^{-3}$
\cite{spergel}, or $R_{g}^{*}=0.613\pm0.028$. The confrontation
of these numbers would, in our approach based on (\ref{eq:SEP-CMB4}),
suggest a rather high negative value for the parameter $\eta_{g}^{*}$.
In other words, assuming that the $^{4}He$ and $^{7}Li$ analyses
really reflect the baryon content of the universe, the gravitational
interaction heavily violates the SEP, at least at the epoch of last
scattering, if the whole discrepancy is accounted for by this effect.
This would be the first experimental evidence that general relativity
is not the correct theory of gravitation. However, large systematic
uncertainties affect the $^{4}He$ and $^{7}Li$ abundance estimation.
These may be related to observation or due to the lack of understanding
of the complex physics in the evolution of these abundances \cite{izotov,ryan,cyburt1}.
Errors and incompletenesses in the standard BBN scheme may also lead
to deviations \cite{cyburt2,coc,cuoco}. Many efforts are made to
reduce these systematic errors.

The deuterium abundance is extremely sensitive to the primordial baryon
content. Moreover it may have been produced in significant quantities
only during the BBN. Its measurement in quasar absorption line systems
is therefore an extremely good probe of the baryon content of our
universe \cite{kirkman,meara,odorico,pettini}. The most recent estimate
of the primordial deuterium-to-hydrogen abundance ratio $D/H$ based
on a recent analysis toward five quasars gives $D/H=2.78_{-0.38}^{+0.44}\times10^{-5}$
\cite{kirkman}. This value corresponds to a weighted average of the
results obtained for each quasar independently. The corresponding
value for the baryon content, in the framework of the standard BBN
theory, reads $\Omega_{b}h^{2}=(21.4\pm2)\times10^{-3}$, or $R^{*(BBN-D)}=0.596\pm0.056$.
Combined with the one-year WMAPext value given here above, this measure
gives the following constraint on a possible SEP violation: $-0.14\leq\eta_{g}^{*}s_{b}^{1*}\leq0.08$.
This bound, once translated into a constraint on the parameter $\eta_{g}^{*}$,
leads to:\begin{eqnarray}
\eta_{g}^{*(BBN-D)} & = & -0.3\pm1\quad.\label{eq:EC-2}\end{eqnarray}
 The determination of the primordial $^{3}He$ abundance is more difficult
as its destruction and production in stars are not well understood.
However a recent upper limit on $^{3}He/H$ leads to a prediction
for the baryon content of the universe in complete agreement with
deuterium measurements \cite{bania}. Remains to be noticed that the
dispersion of the values obtained for the deuterium abundance from
different quasar absorption lines is bigger than expected from individual
measurement errors. This dispersion could be real but the hypothesis
of underestimated systematic errors in the measurements is favored
\cite{kirkman,meara,crighton}. More data would be needed to confirm
the measurements and limit systematics. However, in the framework
of light element abundance measurements, deuterium analysis remains
the most reliable evaluation of the universe's baryon content thanks
to its high sensitivity to the baryon content and the relative absence
of deuterium production after BBN. In this context, the discrepancy
between the baryon content inferred from $D$ and from $^{4}He$ or
$^{7}Li$ analyses should be resolved by a better assessment of the
systematics affecting the measurements of the last two elements abundances.

New physical scenarios beyond the standard BBN are also considered
for solving this apparent tension. Leaving aside the present discrepancies
among the BBN measurements, several proposals have recently been made
for reconciling BBN and CMB measurements. The new physical effects
invoked notably consider the modification of the number of relativistic
particle species, variations of the strength of gravity in the early
universe, or its dependence on the nature of interacting particles
\cite{barger,copi,kneller,masso,barrow,ichikawa}. Our last constraint
on $\eta_{g}^{*}$ may be understood as an alternative solution in
this direction.

\subsection*{Discussion}

First we emphasize that accurate constraints on a primordial SEP violation
should be determined by a best fit of our modified theory with experimental
data. However, the numerical compatibility of the two independent
bounds obtained, (\ref{eq:EC-1}) and (\ref{eq:EC-2}), supports our
results. Also notice that in the framework of a specific alternative
to general relativity the SEP violation is not the only new effect.
The introduction of auxiliary gravitational fields affects the structure
of gravitation itself and notably leaves signatures in the CMB as
well as in the BBN. This will inevitably modify our bounds. In such
a framework, the corresponding bounds on $\eta_{g}^{*}$ could also
be run backward or forward over cosmological timescales for comparison,
either with theoretical predictions on initial conditions ($\eta_{g}^{i}$),
or with present experimental constraints ($\eta_{g}^{0}$).

On the one hand, string theories naturally lead to an effective scalar-tensor
gravity with a running of the parameter $\eta_{g}$ from an initial
value $\eta_{g}^{i}$ of order unity. This initial amplitude of violation
is essentially preserved during the radiation era since the parameter
$\eta_{g}$ depends on the auxiliary scalar field(s) of gravitation,
which is(are) frozen during that period. A large SEP violation at
recombination should therefore be expected in that context. The order
of magnitude of our bounds on $\eta_{g}^{*}$ are still compatible
with such a smooth running of that value until recombination time.
Improved measurements could however rapidly reveal new physics beyond
general relativity.

On the other hand, the experimental constraints at our epoch ($\eta_{g}^{0}\leq1\times10^{-3}$)
require a strong decrease of $\eta_{g}$ between recombination and
today. An attractor mechanism has been advocated for a particular
class of scalar-tensor theories, according to which the scalar coupling
of gravitation, and consequently the parameter $\eta_{g}$, vanish
at late times, to recover general relativity \cite{damour3}. In this
scenario our bounds on the SEP are naturally compatible with the present
experimental limits.

\section{Conclusion}

The SEP is an essential feature of the theory of general relativity,
distinguishing it from any other (experimentally viable) metric theory
of gravitation. A violation of the SEP introduces a cosmological Nordtvedt
effect in the establishment of the acoustic oscillations imprinted
in the CMB temperature power spectrum. The corresponding peaks height
therefore measures a \emph{gravitational baryonic mass density} of
the universe. The modified theory considered here introduces this
effect as the only signature beyond general relativity, orthogonal
to the variation of other cosmological parameters. In this framework
we derived constraints on a possible SEP violation, testing in this
way Einstein's theory of gravitation, through two independent measurements
of the \emph{inertial baryonic mass density} of the universe. The
CMB temperature power spectrum peaks location and the light element
abundances in standard BBN respectively lead to $|\eta_{g}^{*(CMB)}|\leq0.6$
and $\eta_{g}^{*(BBN-D)}=-0.3\pm1$.

More accurate bounds should be determined through a best fit of our
modified theory with the experimental data. We also emphasized that,
in specific alternatives to general relativity, the cosmological Nordtvedt
effect is not the only new effect and the corresponding bounds will
in principle be affected. Finally, our approach also offers a possibility
of understanding apparent discrepancies between CMB and BBN baryon
density measurements in terms of new physics.

\begin{acknowledgments}
The authors wish to thank A. Kosowsky, P. J. E. Peebles and N. Sugiyama
for interesting comments and discussions. The work of V. B. and J.-M.
G. was supported by the Belgian Science Policy through the Interuniversity
Attraction Pole P5/27. Y. W. also acknowledges support of the european
Harmonic Analysis and Statistics for Signal and Image Processing research
network. 
\end{acknowledgments}


\begin{thebibliography}{10}
\bibitem{page}L. Page \emph{et al.}, Astrophys. J. Suppl. Ser. \textbf{148}, 233
(2003). 
\bibitem{spergel}D. N. Spergel \emph{et al.}, Astrophys. J. Suppl. \textbf{148}, 175
(2003). 
\bibitem{bouchet}F. R. Bouchet, Preprint astro-ph/0401108 (2004). 
\bibitem{coles}P. Coles, P. Dinnen, J. Earl, and D. Wright, Preprint astro-ph/0310252
(2003). 
\bibitem{hansen}F. K. Hansen, A. J. Banday, and K. M. G\'{o}rski, Preprint astro-ph/0404206
(2004). 
\bibitem{hajian}A. Hajian and T. Souradeep, Astrophys. J. Lett. \textbf{597}, L5 (2003). 
\bibitem{chen}X. Chen and M. Kamionkowski, Phys. Rev D \textbf{60}, 104036 (1999). 
\bibitem{catena}R. Catena, N. Fornengo, A. Masiero, M. Pietroni, and F. Rosati, Preprint
astro-ph/0403614 (2004). 
\bibitem{nagata2}R. Nagata, T. Chiba, and N. Sugiyama, Phys. Rev. D \textbf{69}, 083512
(2004). 
\bibitem{nagata1}R. Nagata, T. Chiba, and N. Sugiyama, Phys. Rev. D \textbf{66}, 103510
(2002). 
\bibitem{boucher}V. Boucher, J.-M. Gérard, P. Vandergheynst, and Y. Wiaux, Preprint
astro-ph/0407508 (2004). 
\bibitem{gerard}J.-M. Gérard and Y. Wiaux, Phys. Rev. D \textbf{66}, 024040 (2002). 
\bibitem{damour1}T. Damour and G. Esposito-Farèse, Class. Quantum Grav. \textbf{9},
2093 (1992). 
\bibitem{nordtvedt1}K. Nordtvedt, Phys. Rev. \textbf{169}, 1014 (1968). 
\bibitem{nordtvedt2}K. Nordtvedt, Phys. Rev. \textbf{169}, 1017 (1968). 
\bibitem{will}C. M. Will, Living Rev. Rel. \textbf{4}, 4 (2001). 
\bibitem{williams}J. G. Williams, S. G. Turyshev, and T. W. Murphy Jr., Int. J. Mod.
Phys. D \textbf{13}, 567 (2004). 
\bibitem{bailes}M. Bailes, S. M. Ord, H. S. Knight, and A. W. Hotan, Astrophys. J.
Lett. \textbf{595}, L49 (2003). 
\bibitem{kaspi}V. M. Kaspi \emph{et al.}, Astrophys. J. \textbf{543}, 321 (2000). 
\bibitem{esposito}G. Esposito-Farèse, Preprint gr-qc/0402007 (2004). 
\bibitem{hu01}W. Hu, M. Fukugita, M. Zaldarriaga, and M. Tegmark, Astrophys. J.
\textbf{549}, 669 (2001). 
\bibitem{hu96}W. Hu and M. White, Astrophys. J \textbf{471}, 30 (1996). 
\bibitem{hu95a}W. Hu and N. Sugiyama, Phys. Rev. D \textbf{51}, 2599 (1995). 
\bibitem{hu95b}W. Hu and N. Sugiyama, Astrophys. J. \textbf{444}, 489 (1995). 
\bibitem{hu95c}W. Hu, Ph.D. Thesis, UC Berkeley, Preprint astro-ph/9508126 (1995). 
\bibitem{kosowsky}A. Kosowsky, M. Milosavljevic, and R. Jimenez, Phys. Rev. D \textbf{66},
063007 (2002). 
\bibitem{carroll}S. M. Carroll and M. Kaplinghat, Phys. Rev. D \textbf{65}, 063507
(2002). 
\bibitem{damour2}T. Damour and B. Pichon, Phys. Rev. D \textbf{59}, 123502 (1999). 
\bibitem{izotov}Y. I. Izotov and T. X. Thuan, Astrophys. J. \textbf{602}, 200 (2004). 
\bibitem{luridiana}V. Luridiana, A. Peimbert, M. Peimbert, and M. Cervi\~{n}o, Astrophys.
J. \textbf{592}, 846 (2003). 
\bibitem{fields}B. D. Fields and K. A. Olive, Astrophys. J. \textbf{506}, 177 (1998). 
\bibitem{ford}A. Ford \emph{et al.}, Astron. Astrophys. \textbf{393}, 617 (2002). 
\bibitem{Bonifacio}P. Bonifacio \emph{et al.}, Astron. Astrophys. \textbf{390}, 91 (2002). 
\bibitem{ryan}S. G. Ryan, T. C. Beers, K. A. Olive, B. D. Fields, and J. E. Norris,
Astrophys. J. Lett. \textbf{530}, L57 (2000). 
\bibitem{cyburt1}R. H. Cyburt, B. D. Fields, and K. A. Olive, Phys. Lett. B \textbf{567},
227 (2003). 
\bibitem{cyburt2}R. H. Cyburt, Preprint astro-ph/0401091 (2004). 
\bibitem{coc}A. Coc, E. Vangiono-Flam, P. Descouvemont, A. Adahchour, and C. Angulo,
Astrophys. J. \textbf{600}, 544 (2004). 
\bibitem{cuoco}A. Cuoco \emph{et al.}, Preprint astro-ph/0307213 (2004). 
\bibitem{kirkman}D. Kirkman, D. Tytler, N. Suzuki, J. M. O'Meara, and D. Lubin, Astrophys.
J. Suppl. Ser. \textbf{149}, 1 (2003). 
\bibitem{meara}J. M. O'Meara \emph{et al.}, Astrophys. J. \textbf{552}, 718 (2001). 
\bibitem{odorico}S. D'Odorico, M. Dessauges-Zavadsky, and P. Molaro, Astron. Astrophys.
\textbf{368}, L21 (2001). 
\bibitem{pettini}M. Pettini and D. V. Bowen, Astrophys. J. \textbf{560}, 41 (2001). 
\bibitem{bania}T. M. Bania, R. T. Rood, and D. S. Balser, Nature \textbf{415}, 54
(2002). 
\bibitem{crighton}N. H. M. Crighton, J. K. Webb, R. F. Carswell, and K. M. Lanzetta,
Mon. Not. R. Astron. Soc. \textbf{345}, 243 (2003). 
\bibitem{barger}V. Barger, J. P. Kneller, H.-S. Lee, D. Marfatia, and G. Steigman,
Phys. Lett. B \textbf{566}, 8 (2003). 
\bibitem{copi}C. J. Copi, A. N. Davis, and L. M. Krauss, Phys. Rev. Lett. \textbf{92},
171301 (2004). 
\bibitem{kneller}J. P. Kneller and G. Steigman, Phys. Rev. D \textbf{67}, 063501 (2003). 
\bibitem{masso}E. Mass\'{o} and F. Rota, Preprint astro-ph/0406660 (2004). 
\bibitem{barrow}J. D. Barrow and R. J. Scherrer, Preprint astro-ph/0406088 (2004). 
\bibitem{ichikawa}K. Ichikawa, M. Kawasaki, and F. Takahashi, Preprint astro-ph/0402522
(2004). 
\bibitem{damour3}T. Damour and K. Nordtvedt, Phys. Rev. Lett. \textbf{70}, 2217 (1993).
\end{thebibliography}
\end{document}